# An Analysis of Sampling Effect on the Absolute Stability of Discrete-time Bilateral Teleoperation Systems

Amir Aminzadeh Ghavifekr
Faculty of Electrical and Computer Engineering
University of Tabriz
Tabriz, Iran
aa.ghavifekr@tabrizu.ac.ir

Seyedshahab Chehraghi
Faculty of Electrical and Computer Engineering
University of Tabriz
Tabriz, Iran
shahab.chehraghi@ieee.org

Giacomo De Rossi
Department of Computer Science
University of Verona
Verona, Italy
giacomoderossi@outlook.com

*Abstract* - **Absolute stability of discrete-time teleoperation systems can be jeopardized by choosing inappropriate sampling time architecture. A modified structure is presented for the bilateral teleoperation system including continuous-time slave robot, master robot, human operator, and the environment with sampled-data PD-like + dissipation controllers which make the system absolute stable in the presence of the time delay and sampling rates in the communication network. The output position and force signals are quantized with uniform sampling periods. Input-delay approach is used in this paper to convert the sampled-data system to a continuous-time counterpart. The main contribution of this paper is calculating a lower bound on the maximum sampling period as a stability condition. Also, the presented method imposes upper bounds on the damping of robots and notifies the sampling time importance on the transparency and stability of the system. Both simulation and experimental results are performed to show the validity of the proposed conditions and verify the effectiveness of the sampling scheme.**

*Keywords-Teleoperation System; Sampled-data Control; Stability; Transparency; Networked Control Systems; Master-Slave Robots*

## I. INTRODUCTION

Teleoperation systems mostly have been utilized in the remote and hazardous operations such as undersea or space explorations [1, 2], and in delicate applications such as micro-assembly and telesurgery [3]. A throughout review of concepts and principles of bilateral teleoperation mechanisms is studied in [4]. Providing stability and transparency in the presence of the unavoidable communication channel time delay is the main challenging topic for researchers in this area. Several continuous-time control approaches such as passivity theorem [5], wave variable method [6], and adaptive controllers [7] have been utilized to address this issue. All well-known robotic theories such as disturbance rejection methods[8, 9], Lyapunov-based controllers[10], and intelligent control [11] have been extended to teleoperation systems.

Although the numerous amount of studies exist for continuous-time bilateral structures, only a few researches have mentioned stability conditions of the discrete-time bilateral structures[12]. One of the most primary challenges in this area is energy leaking due to the using of the zero order hold devices (ZOHs). Numerous methods have been proposed to overcome this issue. These methods including Tustin approach with the scattering theorem [13, 14], the step invariant mapping with appropriate filters [15], input-state stability concept using nonlinear methods [16], and geometric telemanipulation[17]. In [18], it is assumed that the environment is a virtual wall and the stability conditions are calculated for discrete form of this pattern. Extensions for nonidealities such as quantization, friction, and energy losing in [19] are considered as next steps.

A mathematical method for the stability of the discrete-time teleoperation system has been presented in [20, 21]. The passivity and stability of the delay-free discrete-time teleoperators with position-position architectures have been studied in [22] and [23], respectively. In [23], assuming the accurate dynamics of the ZOHs and ideal samplers, the stability of the system is proved using the small gain theorem. In [24] the effect of the sampling rate on the transparency of the teleoperation system has been studied and the hybrid parameters of the discrete-time system have been calculated. Discrete-time circle criterion has been applied to have absolute stable sampled-data haptic interaction [25]. In this method there is no necessity to have passive operator or environment.

This paper evaluates the influence of the sampling rate on the stability conditions of the sampled-data teleoperators. The position-force architecture is selected which means that the transmitting signals are position of the master and force of the slave robots. The stability conditions impose bounds on the damping parameters of the master and slave, and the sampling rate. These analyses prepare mathematical guidelines to design more transparent and stable bilateral teleoperation systems.

The organization of this paper consists of the following sections: Preliminaries and dynamics of the teleoperation systems are introduced in section II. In section III, proposed discrete-time architecture of bilateral system is presented and the method of finding absolute stability conditions is described. The proposed framework is evaluated for discrete counterpart of the PD-like+dissipation controller in this section to calculate an upper bound for the allowable sampling time without losing the stability. In section IV, the performance of the method has been evaluated by numerical simulations. Finally, an experimental verification is given in section V.

## II. TELEOPERATION MODELLING AND DYNAMICS

In the general architecture of teleoperation system, the master robot which is connected to an operator is moved and its position signals are transmitted through the network to the slave side. The standard scheme of this bilateral form is presented in Fig. 1. It is assumed that the dynamics of the robots are similar.

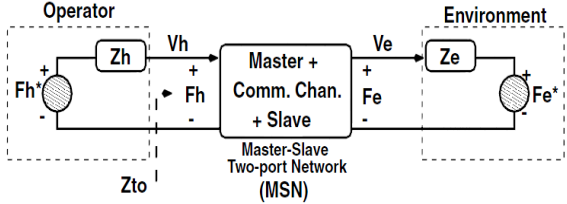

Fig. 1. Standard plot of the bilateral teleoperation system[26]

Assuming master and slave systems as two degree of freedom robots, the well-known dynamics of these systems are written as:

$$m_m \ddot{q}_m + b_m \dot{q}_m = F_h - F_m \quad (1)$$
$$m_s \ddot{q}_s + b_s \dot{q}_s = F_e - F_s$$

where subscripts $s$ and $m$ are implied for the slave and master robots, respectively. The masses and related dampings of robots are given by $m$ and $b$, respectively. $q$ is position state and $\dot{q}$ is used for velocity signals. $F_s$ and $F_m$ are control torques in the slave and master dynamics. Operator and environment dynamics can be described by:

$$F_h = F_h^* - Z_h(s) s X_m \quad (2)$$
$$F_e = F_e^* - Z_e(s) s X_s \quad (3)$$

where $Z_h(s)$ and $Z_e(s)$ denote The LTI impedances of the human operator and the environment, respectively. $F_h^*$ is the exogenous force input applied by the operator and $F_e^*$ is the exogenous force input applied by the environment. To relate the position and force signals of the master and slave sides the so-called Hybrid matrix can be described:

$$\begin{bmatrix} F_h(s) \\ -sX_s(s) \end{bmatrix} = \begin{bmatrix} h_{11} & h_{12} \\ h_{21} & h_{22} \end{bmatrix} \begin{bmatrix} sX_m(s) \\ F_e(s) \end{bmatrix} \quad (4)$$

where

$$h_{11} = Z_m + C_m \frac{Z_s}{Z_s + C_s}, \quad h_{12} = \frac{C_m}{Z_s + C_s}$$
$$h_{21} = -\frac{C_s}{Z_s + C_s}, \quad h_{22} = \frac{1}{Z_s + C_s} \quad (5)$$

where $C_s$ and $C_m$ are local controllers of the slave and master robots, respectively. According to Fig 1, and equations (1-5), it can be deduced that:

$$\frac{X_m}{F_m} = \frac{1}{s} \frac{1}{m_m s + b_m + z_h}, \quad \frac{X_s}{F_s} = \frac{1}{s} \frac{1}{m_s s + b_s + z_e} \quad (6)$$

Applying the dynamics of the zero order holds, following transfer functions can be extracted:

$$G_m(s) = \frac{X_m}{F_m^*} = \frac{1}{s} \frac{1}{m_m s + b_m + z_h(s)} \frac{1 - e^{-sT}}{sT}$$
$$G_s(s) = \frac{X_s}{F_s^*} = \frac{1}{s} \frac{1}{m_s s + b_s + z_e} \frac{1 - e^{-sT}}{sT} \quad (7)$$

Considering the proposed sampling model and equation (5), controllers can be described as:

$$F_m^*(s) = C_m(e^{sT})[-\alpha G_m^*(s) F_m^*(s) + G_s^*(s) F_s^*(s)]$$
$$F_s^*(s) = C_s(e^{sT})[-G_s^*(s) F_s^*(s) + \alpha G_m^*(s) F_m^*(s)] \quad (8)$$

where * as the superscript indicates the discrete-time counterpart of transform functions. $\alpha$ is a scaling factor related to the position signal. The characteristic equation of the sampled-data teleoperation system can be stated as:

$$1 + \alpha C_m(e^{sT}) G_m^*(s) + C_s(e^{sT}) G_s^*(s) \quad (9)$$

In [23], using the small gain theorem, the absolute stability of the aforementioned closed loop system is proved. It is declared that the position error-based teleoperation system is stable if and only if satisfies the following inequality:

$$\|M_m N_m + M_s N_s\|_\infty < 1 \quad (10)$$

where

$$N_m = \frac{\alpha b_s C_m(e^{sT}) r(s)}{2 b_m b_s + \alpha b_s C_m(e^{sT}) r(s) + b_m C_s(e^{sT}) r(s)}$$

$$N_s = \frac{b_m C_s(e^{sT}) r(s)}{2 b_m b_s + \alpha b_s C_m(e^{sT}) r(s) + b_m C_s(e^{sT}) r(s)}$$

$$M_m = -1 + \frac{2 b_m}{r(s)} G_m^*(s), \quad M_s = -1 + \frac{2 b_s}{r(s)} G_s^*(s) \quad (11)$$

$$r(j\omega) = \frac{T}{2} \frac{e^{-j\omega T} - 1}{1 - \cos \omega T}$$

where $T_1$ and $T_2$ are forward and backward delays which are integer multiple of sampling period. Assuming $\alpha = 0$, the aforementioned stability inequality can be described by:

$$\frac{|D + b_s C_m r| + |D + b_m C_s r| + |D|}{|2 b_m b_s C_m C_s + b_s C_m^2 C_s r + b_m C_s^2 C_m r + D|} < 1 \quad (12)$$

In which

$$D = \frac{r^2 (1 - e^{-(T_1 + T_2)s})}{2} \quad (13)$$

## III. THE PROPOSED SAMPLED-DATA ARCHITECTURE OF BILATERAL TELEOPERATION SYSTEM

The proposed mathematical structure of the sampled-data teleoperation is presented in this section.

The sampling rate for output signals of both robots assumed to be equal and is denoted by $\hat{t}_k$, $k \in N$. All sampled position and velocity signals are sent in form of data packets via communication channel which suffers from constant time delay. The samplers in this scheme are time driven while the two zero order holds are event driven. The proposed model is presented in Fig. 2, so that the dash lines are used to illustrate the sampled signals.

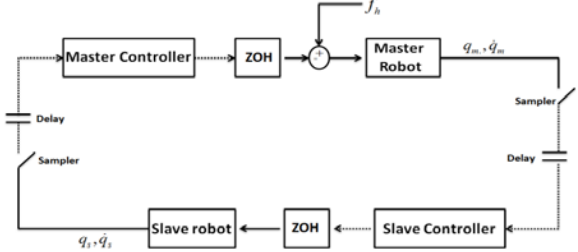

Fig.2. The general structure of the discrete-time teleoperation

It is presumed that the velocity and position signals are sampled at $\hat{t}_k$. Also, a constant time delay $T$ is applied to state signals during the network transmission. This delay can be larger than interval $[\hat{t}_k, \hat{t}_{k+1}]$. Duration of the $k$th sampling period is calculated by $h_k$, i.e. $h_k = \hat{t}_{k+1} - \hat{t}_k$

**Assumption 1.** There exists $\varepsilon > 0$ such that $\hat{t}_{k+1} - \hat{t}_k > \varepsilon$. This assumption notifies that sampling intervals cannot perform simultaneously in practical system.

Update rates of the zero order holds at the instants $t_k$ are:
$$t_k = \hat{t}_k + T_k \quad k \in N \tag{14}$$

The duration of last sampling constant $\hat{t}_k$ is calculated as:
$$\mu(t) \triangleq t - \hat{t}_k = t - t_k + T \tag{15}$$

$\mu(t)$ is defined as the induced delay. The maximum amount of this network-induced delay represented by $\gamma$ is calculated as:
$$\gamma = \sup(\mu(t)) = \sup(\hat{t}_{k+1} - \hat{t}_k) \tag{16}$$

The utilized controllers for master and slave robots in the proposed structure of Fig. 2 are PD like controllers + dissipations. An emulated mode of this controller is proposed in [27] to improve the accuracy of the force tracking and achieve better coordination of robots. The continuous-time forms of controllers are proposed as:
$$\tau_1(t) = -K_v(\dot{q}_1(t) - \dot{q}_2(t - \tau_2)) - (K_d + P_\varepsilon)\dot{q}_1(t) - K_P(q_1(t) - q_2(t - \tau_2))$$
$$\tau_2(t) = -K_v(\dot{q}_2(t) - \dot{q}_1(t - \tau_1)) - (K_d + P_\varepsilon)\dot{q}_2(t) - K_P(q_2(t) - q_1(t - \tau_1)) \tag{17}$$

where $T_1, T_2 \geq 0$ are delays from master to slave and vice versa. $K_v, K_p$ are the symmetric and positive definite gains, $K_d$ is the positive dissipation gain and $P_\varepsilon$ is an extra damping to protect master-slave coordination. It is proofed in [27], that choosing $K_d = \frac{\nu}{2} K_p$ where $\nu > 0$ is an upper bound of the general delay $T_1 + T_2$ leads to have a passive teleoperation system. Also, if the operator and the environment are passive, the position tracking error between robots will be bounded. Furthermore, if the velocity and accelerations signals converge to the zero, force tracking error will be achieved.

The primary discrete-time form of control signals in (17) can be rewritten as:
$$F_m(t) = -K_v(\dot{x}_m(\hat{t}_k) - \dot{x}_s(\hat{t}_k - T_2)) - (K_d + P_\varepsilon)\dot{x}_m(\hat{t}_k) - K_P(x_m(\hat{t}_k) - x_s(\hat{t}_k - T_2))$$
$$F_s(t) = -K_v(\dot{x}_s(\hat{t}_k) - \dot{x}_m(\hat{t}_k - T_1)) - (K_d + P_\varepsilon)\dot{x}_s(\hat{t}_k) - K_P(x_s(\hat{t}_k) - x_m(\hat{t}_k - T_1)) \tag{18}$$

It is notable that the gains of controllers for slave and master robots are equal. This is due to the similar dynamics of these robots.

The input-delay method is proposed in [28] to calculate the maximum allowable network delay which help to preserve the exponential stability of the discrete-time systems. Using this

Approach, (18) can be rewritten as:
$$F_m(t) = -K_v(\dot{x}_m(t - \mu_s) - \dot{x}_s(t - \mu_s - T_2)) - (K_d + P_\varepsilon)\dot{x}_m(t - \mu_s) - K_P(x_m(t - \mu_s) - x_s(t - \mu_s - T_2))$$
$$F_s(t) = -K_v(\dot{x}_2(t - \mu_s) - \dot{x}_m(t - \mu_s - T_1)) - (K_d + P_\varepsilon)\dot{x}_s(t - \mu_s) - K_P(x_s(t - \mu_s) - x_m(t - \mu_s - T_1)) \tag{19}$$

By substituting the controllers (19) in the absolute stability condition of (11) and using bilinear transformation method we have:
$$\tau_1(t) = \tau_2(t) = -K_v \frac{z-1}{Tz} - (K_d + P_\varepsilon)\frac{z-1}{Tz} - K_P \tag{20}$$

Thus, the stability condition can be simplified to:
$$b_m, b_s > K_p T + 2K_d - 2P_\varepsilon - 2K_v \tag{21}$$

## IV. NUMERICAL SIMULATION RESULTS

The proposed stability conditions have been tested on the 1-DOF teleoperation system modeled by the mass and damping terms. By simulation, the effect of sampling rate on the behavior of sampled-data teleoperation system has been studied. The gains for the PD-like with dissipation controller are chosen $K_p = 1, K_d = 2, P_\varepsilon = 0.002, K_v = 10$.

The environment acts like a stiff wall, which reflecting the overall torque of the slave robot. The following scenario has been considered in simulation. A step force has been applied to the master robot by the human operator for 10 seconds from 10s till 20s. The slave robot meets the environment at $4 rad$. The spring-mass structure is used to model the human operator. Spring coefficient is $10 N/m$ and damping gain is chosen $1 Ns/m$. The generated force by the operator is illustrated in Fig.3

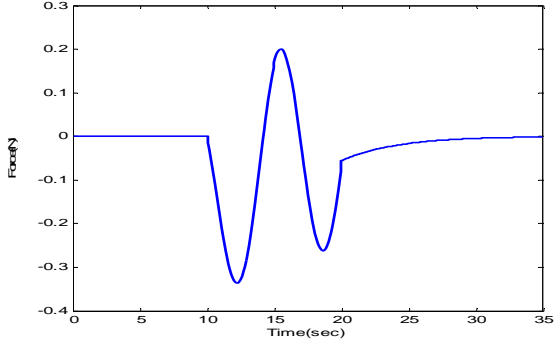

Fig. 3. External force generated by the operator

Both master and slave systems are assumed as one degree of freedom robots with transform function of $M(s) = 2/(2+s)$. Fig. 4 and Fig. 5 present the position and force signals of the master and slave robots, respectively. The maximum allowed sampling period is chosen 0.006s according to (21).

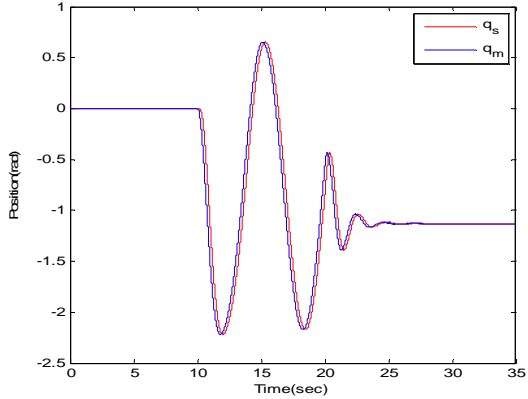

Fig. 4. Master and slave position signals for sampled-data counterpart of PD-like+dissipation controller

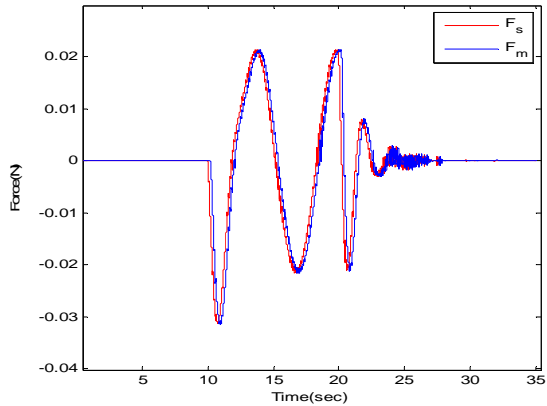

Fig. 5. Master and slave force signals for sampled-data counterpart of PD-like+dissipation controller

The most noticeable remark is the stability can be jeopardized by increasing gains of controllers and sampling time. Also, the physical and practical parameters of the robot cannot be varied frequently. Thus, there should be a trade-off between sampling rate and controller characteristics.

## V. EXPERIMENTAL SETUP AND RESULTS

The system has been tested on a physical setup to establish the validity of the simulation's results. The setup is two brushless DC motors attached to two gearboxes for torque multiplying and two handles. One degree of freedom force sensor is attached to each handle to measure the amount of force/torque of the user and the environment exert on the system. The motor actuators are connected via EtherCAT to a PC, running a soft real-time Linux kernel. This platform provides high frequency software control and allows testing different teleoperation configurations, such as position-position (P-P), force-position (F-P), and 4-channel architecture. By using native kernel libraries and C++ programming, it is possible to guarantee high performances given any desired controller for the system.

Working on a physical setup introduces limitations in both the controller output and the measurement obtainable by the sensors that are usually neglected in a simulation. For instance, the motors accept voltage control in the range of $\pm 5V$ and both position and force sensors produce quantized and noisy signals that require filtering. Additionally, the register used to store the encoder value overflows after approximately fifty full rotations. Therefore, testing a system in simulation and physical setup can be instructive to see whether the sampling time condition or the passivity conditions still hold.

The simulation has been enriched by introducing the parameters resulting from the setup identification process. The latter involved the extraction of data from the motor datasheet and the linear regression over acquired samples from free-runs of the system. These parameters are: $50\,Hz$ for the first-order velocity filters cutoff frequency, according to the spectral analysis of the raw data; $4.054\,V/N$ force-to-voltage coefficient (from motor and force sensor datasheet comparison); $0.1\,m$ arm length for the lever attached to the motor; $2\pi/4096\,rad/step$ quantization for the step encoder (from datasheet). The noise of the sensor has been represented as white Gaussian noise $(0\mu, 1\sigma)$. Finally, the motor, with the attached gearbox and lever, has been identified by using a Kalman smoother on the velocity signal. The result is the following first-order continuous-time transfer function: $M(s) = 19.34/(1.217s+1)$, with rotor inertia $J = 23.54\,kgm^2$ and viscous friction coefficient $F = 0.0517$.

The general scheme of the system is depicted in Figure 6. The first part represents the entire system including master and slave, while the second part shows the details of the master manipulator.

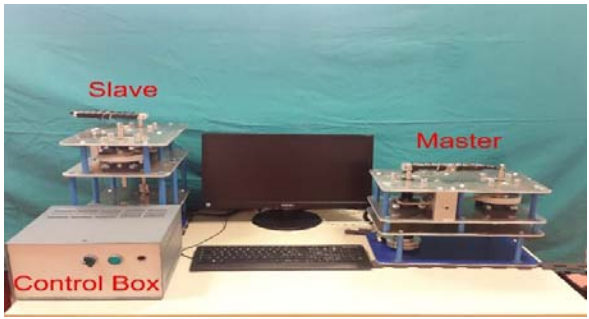
(a)

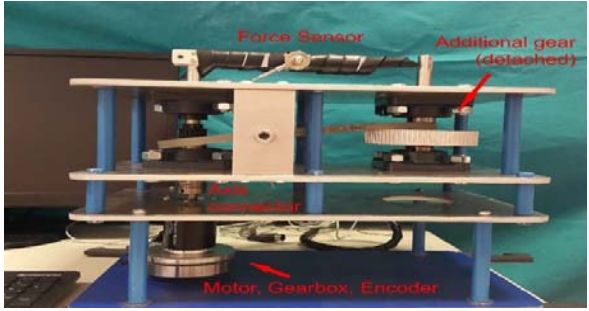
(b)

Fig. 6. a) General scheme of the experimental system b) Components of the master manipulator

The configuration used in this experiment is a stable one with $Kp = 8.4$ and $Kd = 0.0005$. The environment is placed approximately at $3.5\, rad$, as in the simulation. The evaluated maximum sampling time for stability, obtained by the evaluation of equation (21), is $T = 0.006\, s$; due to internal time delays in the control loop, such value decreases down to $T = 0.003\, s$. When this limit is reached, the system presents an unstable behavior.

The position and force tracking signals, presented in Figures 7 and 8, show a scenario in which the slave suddenly hits the obstacle with three different velocities. In the first two contacts, the controllers are quick to stabilize the motion; in the third contact, the greater impact force carried by the increased momentum results in a finite oscillation, which finally stops near $2.5\, rad$.

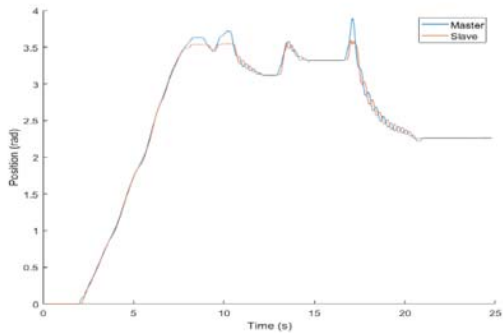
Fig. 7. Master and slave robots position signals for experimental discrete-time PD-like+ dissipation controller

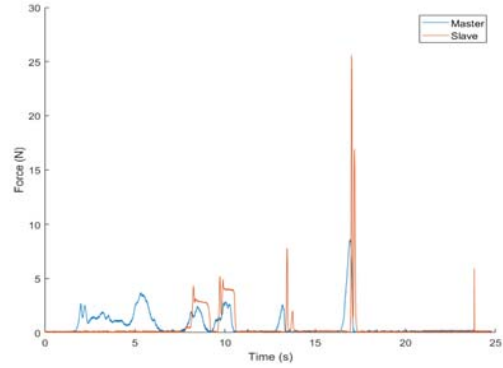
Fig. 8. Master and slave robots force signals for experimental discrete-time PD-like+ dissipation controller

## VI. CONCLUSION

Despite extensive researches about continuous-time teleoperation systems, the sampled-data structures have not been studied widely in the control literature. The stability condition of these systems depends on the rate of the sampling time and unavoidable delay of the communication channel. In this paper, according to the proposed method, limitations of the passivity conditions are omitted due to the non-passive equations of slave and master dynamics and arbitrary passive models of the operator and environment. It is noticeable that although selecting larger controller gains can provide better transparency in the continuous-time structures, there should be a trade-off between the stability conditions and transparency of the system in the discrete-time structures. As a suggestion for future studies on the discrete-time teleoperators, the variable time delay of the network and the quantization error effects of the samplers can be taken into account. Also, the proposed stability conditions can be extended to the 4-channel architecture of the bilateral teleoperation systems.